\documentclass[traditabstract]{aa}
\usepackage{graphicx}
\usepackage{enumerate}
\usepackage{epsfig}
\usepackage{color}
\usepackage{txfonts}
\usepackage{natbib}
\bibpunct{(}{)}{;}{a}{}{,} 

\begin{document}

\title{Why the Milky Way's bulge is not only a bar formed from a cold
thin disk}
\titlerunning{Why the Milky Way's bulge is not only a bar formed from a cold
thin disk}

\author{P. Di Matteo\inst{1}, A. G$\rm \acute o$mez\inst{1},  M. Haywood\inst{1},  F. Combes\inst{2},  M.~D. Lehnert\inst{3,4}, M. Ness\inst{5}, O.~N. Snaith\inst{6}, D. Katz\inst{1},  B. Semelin\inst{2,4}}

\authorrunning{Di Matteo et al.}

\institute{GEPI, Observatoire de Paris, CNRS, Universit\'e
  Paris Diderot, 5 place Jules Janssen, 92190 Meudon, France\\
\email{paola.dimatteo@obspm.fr}
\and
LERMA, Observatoire de Paris, CNRS, 61 Av. de l$'$Observatoire, 75014 Paris, France
\and Institut d{'}Astrophysique de Paris, UMR 7095, CNRS,  98 bis boulevard Arago, 75014 Paris, France
\and Universit$\rm \acute{e}$ Pierre et Marie Curie, 4 place Jussieu, 75005 Paris, France 
\and Max-Planck-Institut f$\rm \ddot{u}$r Astronomie, K$\rm \ddot o$nigstuhl 17, D-69117 Heidelberg, Germany 
\and Department of Physics \& Astronomy, University of Alabama, Tuscaloosa, Alabama, USA
}

\date{Accepted, Received}

\abstract{By analyzing a N-body simulation of a bulge formed simply via a bar instability mechanism operating on a kinematically cold stellar disk, and by comparing the results of this analysis with the structural and kinematic properties of the main stellar populations of the Milky Way bulge, we conclude that the bulge of our Galaxy is not a pure stellar bar formed from a pre-existing thin stellar disk, as some studies have recently suggested. On the basis of several  arguments emphasized in this paper, we propose that the bulge population which, in the Milky Way, is observed not to be part of the peanut structure corresponds to the old galactic thick disk, thus implying that the Milky Way is a pure thin+thick disk galaxy, with only a possible limited contribution of a classical bulge.  }


\maketitle

\section{Introduction}

The inner regions of the Milky Way (hereafter MW) keep traces of the early phases of formation of the Galaxy, and of its subsequent evolution.\\
From atmospheric chemical abundance studies, we know that $\alpha-$enhanced, metal-poor stars are part of the bulge -- the prominent out-of the plane structure characterizing the inner few kpcs of the Galaxy-- and this has been interpreted as evidence of its early and rapid enrichment history \citep{mcwilliam94, zoccali06, fulbright07, lecureur07, mcwilliam08}. However, chemical evolution studies also reveal that some of the stars currently  in the bulge must have formed in a slower and more quiescent star formation episode, as indicated by the presence of  not $\alpha-$enhanced, more metal-rich stars \citep{bensby11, hill11}, preferentially found closer to the Galactic plane \citep{ness13a}.\\
From stellar kinematics studies, we learn that the velocity ellipsoid of moderate metal-rich ([Fe/H] $>$ -0.5 dex), not $\alpha-$enhanced stars, shows a vertex deviation consistent with those stars supporting a bar-like structure \citep{zhao94, soto07, babusiaux10}. However, such studies also reveal that the most metal-poor, $\alpha-$enhanced stars are not part of this elongated structure, having velocity dispersions consistent with a kinematically hotter component \citep{babusiaux10, uttenthaler12, ness13b}.\\
Thus, observational studies suggest that the MW's bulge does
not consist of a single unique component, but is rather a combination
of two or more components or perhaps a continuum of populations, with
different chemical and kinematic properties.
Even if there is no consensus yet on the origin of the $\alpha-$enhanced, metal-poor stars -- classical bulge/old spheroid \citep{babusiaux10, gonzalez11, hill11, uttenthaler12, zoccali14} or thick disk \citep{ness13b, dimatteo14} -- most of the above cited studies seem to agree on the thin disk origin of the not $\alpha-$enhanced, metal-rich component and its current structure: a bar that went through one or multiple vertical instability events in the past,  which led to its current thick, boxy shape, as described by N-body models \citep[see, for example, ][]{combes81, athanassoula05, debattista06, martinez06, ness12, dimatteo14}. \\

If observational studies suggest a complex scenario for the formation of the MW's bulge, with the co-existence of multiple or a continuum of components, possibly formed at different times of the Galactic evolution, several N-body models have suggested that a pure thin disk instability model can explain  sufficiently well the observed
characteristics, without the need to add any significant kinematically hot component -- classical bulge or thick disk \citep{shen10, martinez11, kunder12, martinez13, vasquez13, gardner14, zoccali14}. 
For example, \citet{shen10} analyzed a N-body simulation of a disk galaxy which developed a boxy/peanut-shaped bulge and compared it to the stellar kinematics of the bulge region,  from the  BRAVA bulge survey, concluding that "the model fits the BRAVA stellar kinematic data covering the whole bulge strikingly well with no need for a merger-made classical bulge". \citet{martinez13} showed that pure bar instability models are also able to reproduce vertical metallicity gradients in the Galactic bulge, similar to those observed, provided that the initial disk had a steep enough radial metallicity gradient. They were also able to produce a longitude-latitude (hereafter ($l,b$)) metallicity map remarkably similar to that   constructed by \citet{gonzalez13} from the VVV survey, thus highlighting the result that "a simple model for the Milky Way's boxy bulge", through disk instability, is able to reproduce many of the characteristics observed in the Galactic bulge. Finally, \citet{zoccali14} used the N-body model presented in \citet{martinez11} and compared it with the bulge kinematics from the  GIBS survey, noting that no additional kinematically hotter component needs to be added, at any latitude, to the boxy/peanut-shaped bulge to reproduce the rotation curve and velocity dispersions profiles, thus remarking that  "The very good agreement between this model and the data supports the conclusion presented in Shen et al. (2010)".  However, note also that in the same work, \citet{zoccali14} cautioned the reader about interpreting the MW's bulge as the result of a pure secular evolution process, recalling, among others, the observational evidence given at the beginning of this Section, about the composite nature of the Galactic bulge. \\To conclude, none of the afore-mentioned N-body models pointed out  the necessity to add an additional, kinematically hotter, component to the boxy/peanut-shaped structure, because such a simple scenario was shown to be already able to reproduce all the characteristics considered.\\

Thus, there is clearly a \emph{tension between observations} and their interpretation. On the one side, pure bar instability models fit fairly well the global  trends observed in the MW's bulge, without the need to invoke the presence of any additional component -- and in some cases, clearly excluding it \citep{shen10} on the basis of this good match. On the other side observations present a much more complex scenario, where the Galactic bulge consists of the co-existence of different populations, with different kinematic and chemical properties.\\ 

The aim of this paper is  to alleviate this tension, by  discussing where and why a pure bar instability mechanism operating on a thin disk fails in representing the complexity of the MW's bulge.  By means of a high resolution
N-body simulation of a thin disk which has undergone a bar instability,
we will show that this scenario, despite explaining successfully a number
of the properties of the bulge, is indeed insufficient when the detailed
properties of its main populations are taken into account.
We will show in particular that it is true that N-body simulations which suggest a pure thin disk origin for the MW's bulge can reproduce: \emph{a)} the velocity rotation curve and the velocity dispersion profiles of stars in the bulge at different latitudes, as suggested  by \citet{shen10, kunder12, zoccali14}; \emph{b)} the vertical metallicity gradient and $(l,b)$-metallicity maps of the bulge, qualitatively similar to observations, as proposed by \citet{martinez13} for appropriate initial conditions in the disk, \emph{but} that these constraints alone are not sufficient to validate a pure bar instability mechanism at the origin of the MW's bulge. 
Such models lead indeed to bulges with properties not compatible with those observed for the MW.\\ If the MW's bulge had uniquely a thin disk origin and it was only the result of a pure bar instability mechanism originating in a thin stellar disk with a steep enough initial radial metallicity gradient,  all bulge stars (from the most metal rich to the most metal poor ones) should be part of the boxy/peanut-shaped structure. We will show that, if this was the case: 
\begin{enumerate} 
\item all red clump stars in the MW's bulge with [Fe/H]  $>$ -1 dex should show a split in the distribution of their $K$ magnitudes, which is not observed \citep{ness13a};  \item the metal-poor population (-1 dex $<$ [Fe/H] $\le$ -0.5 dex) should be a kinematically  warm replica of the more metal rich ones (-0.5 dex $<$ [Fe/H]), which is not  \citep{ness13b}. 
\end{enumerate}
New observational results that combine the elemental
abundances and the kinematics allow us to reassess the evolutionary
scenario for the bulge of the MW, and lead us to \emph{exclude that the MW's bulge has a pure thin disk origin. This structure is not simply a thick, boxy/peanut-shaped bar formed in a kinematically cold stellar disk and seen edge-on}. Its most metal-poor (-1 dex $<$ [Fe/H] $\le$-0.5 dex) population does not have a thin disk origin, as recently suggested \citep{martinez13}, but it is rather associated to an ab-initio kinematically warmer component that -- on the basis of several arguments recalled in this paper -- we associate to the old Galactic thick disk.

\section{Simulations}\label{sim}

The simulation analyzed in this paper is one of a set of three high resolution simulations, with varying bulge--to--disk ratios, already  described and analyzed in \citet{dimatteo13, dimatteo14}. It consists of an isolated stellar disk, with a B/D=0.1 classical bulge\footnote{In the following, by classical bulge we mean a spheroidal com- ponent, not formed by disk instabilities, but rather through mergers or some dissipative collapse at early phases of the galaxy formation.}, and containing no gas. We have chosen to present in this paper the results for the simulation with B/D=0.1 because this ratio is at the suggested upper limit of any classical bulge in the MW \citep{shen10, kunder12, dimatteo14}. However, we emphasize that the results obtained for the case with B/D=0.1 are identical to those with B/D=0, making the two scenarios (pure thin disk \emph{versus} thin disk + small classical bulge) \emph{de facto} indistinguishable in the context of the present study. An example of this strong similarity can be appreciated comparing Figs.~9 and 12 of \citet{dimatteo14}: adding a B/D=0.1 classical bulge to the simulation has no impact on the global kinematic characteristics, neither on the velocity curve nor on the velocity dispersion profiles of the boxy bulge. \\
The dark halo and the bulge are modeled as Plummer spheres \citep{BT87}. The dark halo has a mass $M_H = 1.02 \times 10^{11} M_{\odot}$ and a characteristic radius $r_H=10$~kpc. The bulge has a mass $M_B = 9 \times 10^9M_{\odot}$ and characteristic radius $r_B=1.3$~ kpc. The stellar disk follows a Miyamoto-Nagai density profile \citep{BT87}, with mass $M_* = 9 \times 10^{10} M_{\odot}$ and vertical and	radial	scale	lengths given by $h_* =0.5$~kpc	and	$a_* =4$~kpc, respectively. The initial disk size is 13 kpc, and the Toomre parameter is set equal to Q=1.8. The galaxy is represented by $N_{tot}$= 30720000 particles redistributed among dark matter ($N_H$= 10240000) and stars ($N_{stars}$ = 20480000). To initialize particle velocities, we adopted the method described in \citet{hern93}. A Tree-SPH code \citep{sem02} has been used to run the simulations.  A Plummer potential is used to soften gravity at scales smaller than $\epsilon$ = 50 pc. The equations of motion are integrated over 4 Gyr, using a leapfrog algorithm with a fixed time step of $\Delta t = 2.5 \times 10^4$ yr.

\begin{figure}
\centering
\includegraphics[trim = 3cm 7cm 2cm 9cm, clip=TRUE,width=0.5\textwidth,angle=0]{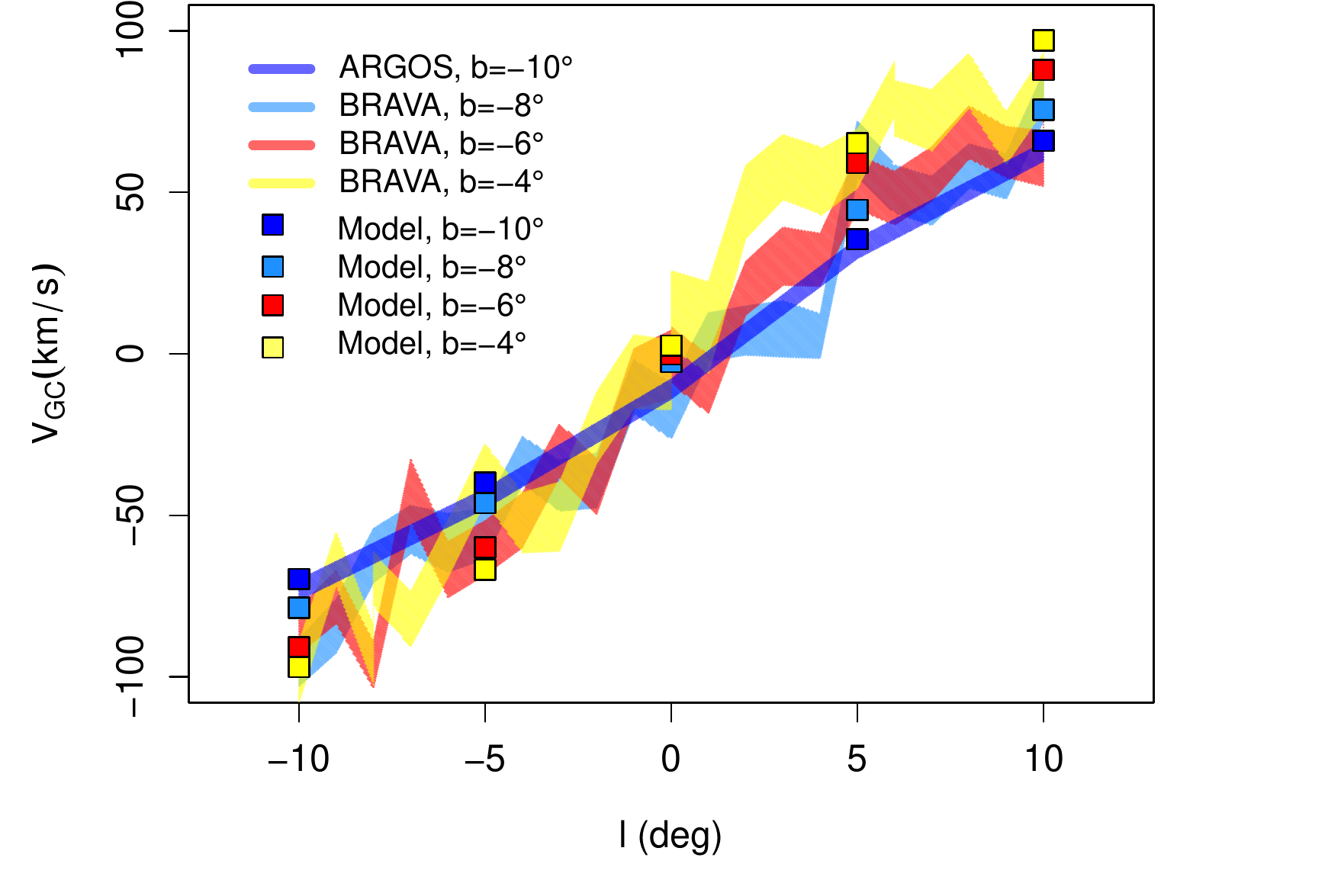}
\includegraphics[trim = 3cm 7cm 2cm 9cm, clip=TRUE,width=0.5\textwidth,angle=0]{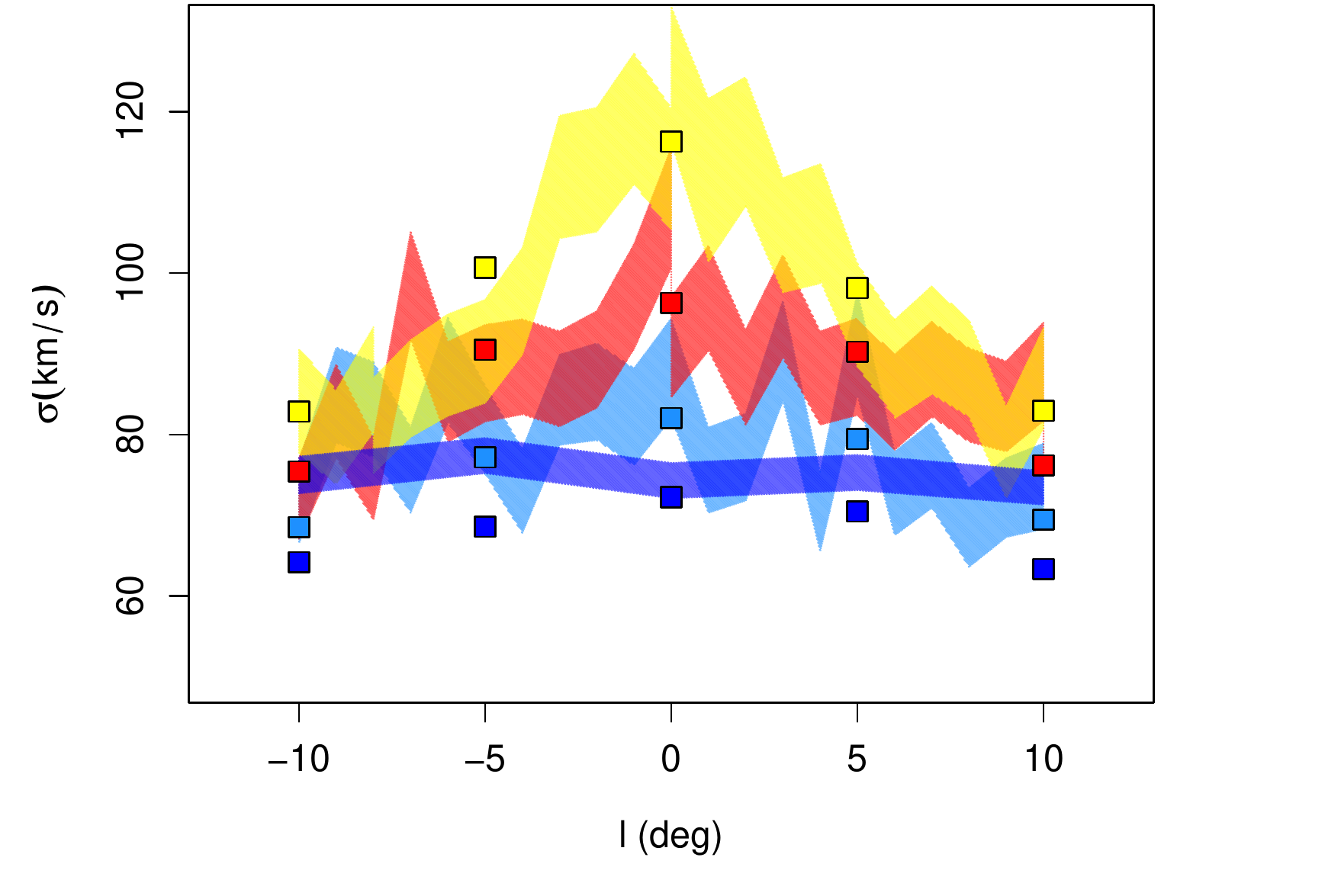}
\caption{Rotation curve \emph{(top panel)} and velocity dispersions \emph{(bottom panel)} of stars of the N-body models at $|x|\le 2.5$~kpc and $|y|\le 3$~kpc from the galaxy center.  Four different latitudes are shown for the modeled galaxy: $b=-4^\circ$ (yellow squares), $b=-6^\circ$ (red squares), $b=-8^\circ$ (pale blue squares), $b=-10^\circ$ (dark blue squares). For comparison, BRAVA fields at  $b=-4^\circ$ (yellow curve), $b=-6^\circ$ (red curve), $b=-8^\circ$ (pale blue curve), and ARGOS fields at  $b=-10^\circ$ (dark blue curve) are also given. The thickness of the curves corresponds to the $\pm 1 \sigma$ error in the observational data.}
\label{totkin}
\end{figure}

\section{Results}

In what follows, the N-body model described in Sect.~\ref{sim} and extensively studied in \citet{dimatteo13, dimatteo14} has been rescaled to match the MW bar size and bulge velocities\footnote{Note that the rescaling is possible because gas is not included in the simulations.}. To this aim, we have divided the positions by a factor about 2, to have a bar length of about 3.5 kpc, the corotation and OLR at 4.5~kpc and 7.5~kpc respectively. With this rescaling, the Sun position is (0., -8., 0.)~kpc. Velocities have been divided accordingly by a factor $\sqrt{2}$, so to match the rotation curve of the MW's bulge, as deduced by radial velocities measurements \citep[see][and  Fig.~\ref{totkin}]{kunder12, ness13b}. The bar is then observed at an angle of 20 degrees with respect to the Sun-Galaxy center direction \citep[see for example][]{bissantz02}. \\

\subsection{Successes of a pure thin-disk bar instability scenario}
The rotation curve and velocity dispersion, as deduced by radial velocity measurements, of the final model are shown in Fig.~\ref{totkin}, together with data from the ARGOS survey for the fields at $b=-10^\circ$ \citep[see][]{ness13b} and from the BRAVA survey for the fields at $b=-4^\circ, -6^\circ, -8^\circ$ \citep{kunder12} . The simulation clearly reproduces the overall kinematic trends  found for the MW's bulge, namely the nearly independence of the rotation curve with latitude\footnote{Note that a mild velocity gradient with latitude is found for stars at positive or negative longitudes, that is with $|l| > 0$. For example, at $l=-10^\circ$, the velocity gradient is of about 6~km/s/deg, stars at $b=-10^\circ$ having a rotational velocity which is about 35~km/s smaller than those of stars at   $b=-4^\circ$. A similar behavior is found also in observational data, see for example \citet{ness13b}.  } (i.e. cylindrical rotation) and the decrease and flattening of the velocity dispersions with vertical distance from the plane. Thus, in agreement with previous results \citep{shen10, kunder12, gardner14, zoccali14},  this N-body model supports the finding that a pure bar instability scenario operating on a thin disk can fit the \emph{overall} kinematics of the MW's bulge, without the need of adding a massive kinematically hotter component (a classical bulge with B/D$>$0.1 or a thick disk) to explain the observed global kinematic trends.

\begin{figure}
\centering
\includegraphics[trim = 0cm 0cm 0cm 0cm, clip=TRUE,width=0.35\textwidth,angle=0]{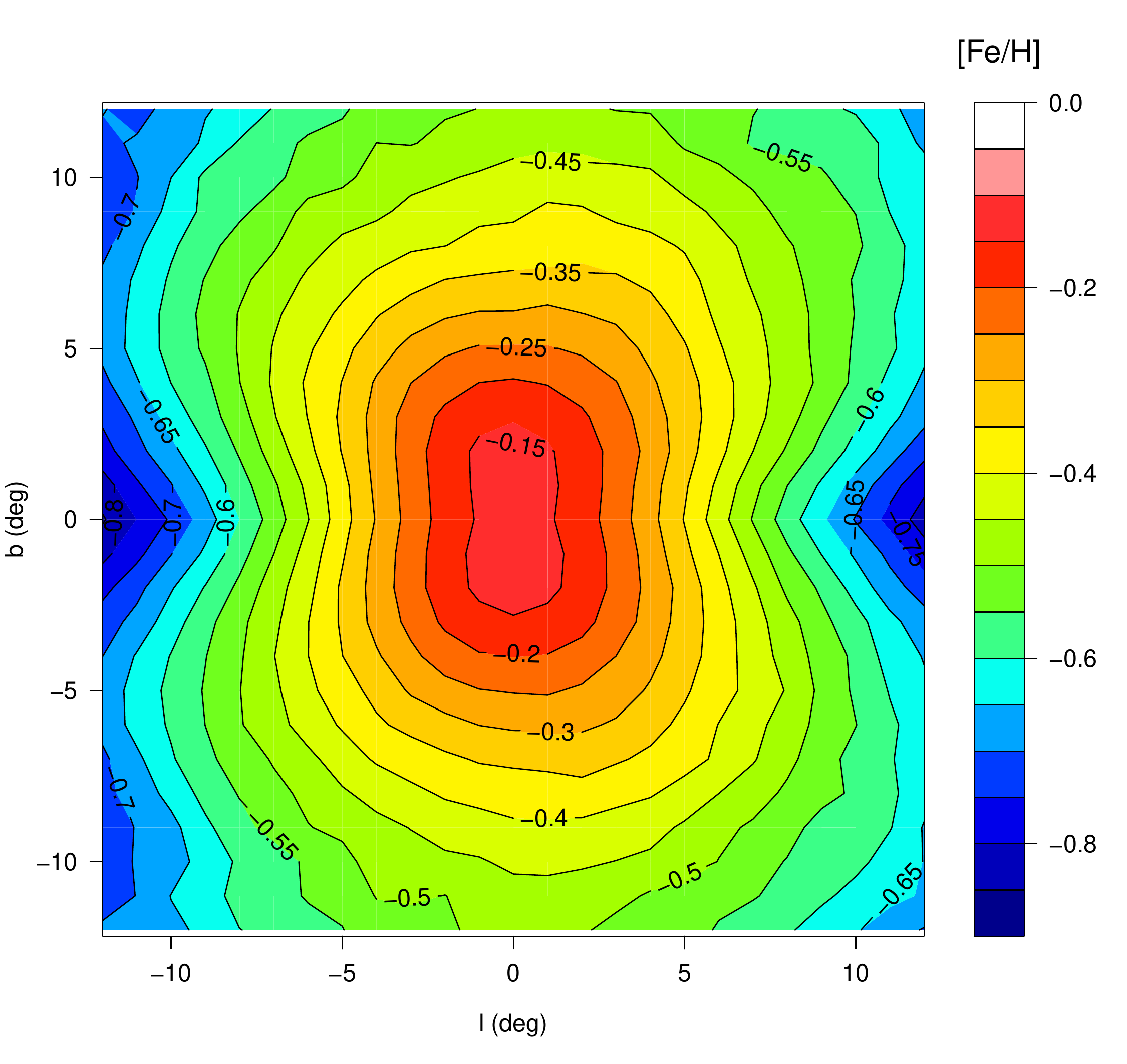}
\includegraphics[trim = 0cm 0cm 0cm 0cm, clip=TRUE,width=0.35\textwidth,angle=0]{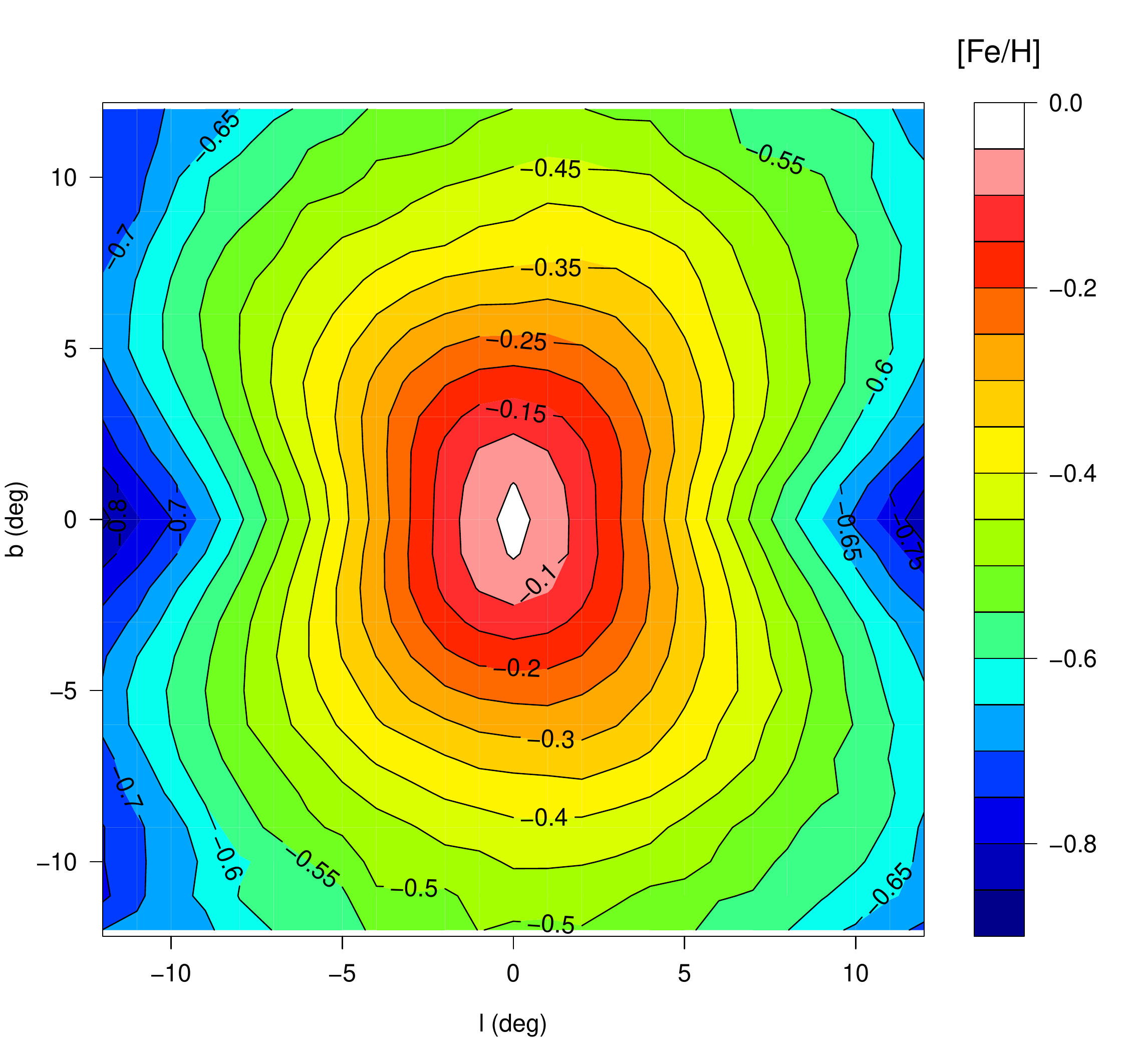}
\caption{Metallicity maps of the simulated galaxy in galactic coordinates: \emph{Top panel:} Disk as well as classical bulge stars are included; \emph{Bottom panel:} Disk stars only are shown (see text for the choice of the initial metallicity of disk and classical bulge stars). In both panels, only stars with $|x|\le 2.5$~kpc and $|y|\le 3$~kpc from the galaxy center have been selected. Some contour levels for the metallicity maps are also reported, with values given in the maps.  }
\label{metmap}
\end{figure}

Thin disk instability scenarios for the formation of the MW's bulge have been shown to be successful also in generating vertical metallicity gradients and metallicity maps consisting with observations. In particular, \citet{martinez13} have shown that  stars in a bar can keep memory of their initial conditions in such a way that an initial (i.e. before bar formation) radial metallicity gradient in the thin stellar disk can be translated into a vertical gradient in the bulge. An initial radial gradient in the thin disk  sufficiently steep\footnote{Note that the pure thin disk scenario requires that the initial radial metallicity gradient  is steep at all radii. The adoption of a shallower gradient at radii greater than the bulge size, as done by \citet{bekki11b},  can lead to a final vertical gradient significantly weaker than that observed in the MW bulge. We refer the reader to the discussion in \citet{martinez13} on this point.} ($\sim -0.4$~dex/kpc, as in \citet{martinez13}) would reflect into a vertical gradient along the bulge minor axis similar to that observed \citep{zoccali08, gonzalez11, gonzalez13}.
In Fig.~\ref{metmap}, we present the $(l,b)-$mean metallicity maps of our simulated  galaxy.  The initial radial metallicity  profile assigned to the thin stellar disk before bar formation is $\rm{[Fe/H]}=\rm{[Fe/H]_0}+\alpha R$, with $\alpha= -0.4$dex/kpc, $R$ the distance in the plane from the galaxy center, and $\rm{[Fe/H]_0}= 0.5$~dex being the metallicity at $R=0$. For the classical bulge, we adopt a gaussian metallicity distribution, with a mean at -0.4~dex, which is the typical metallicity of a spheroid of that mass \citep[see][]{gallazzi05, lee08, thomas10}, and a dispersion of 0.3~dex. Similarly to \citet{martinez13}: \emph{1)} the metallicity maps show a characteristic peanut shape, with the major axis of the peanut elongated parallel to the minor axis of the stellar bar; \emph{2)}  a vertical gradient (of the order of $\sim -0.04~\rm{dex/deg}$ in our model, similar to the estimates given in \citet{gonzalez13}) can be generated in the bulge. These findings are a consequence of the mapping of thin disk stars into the bulge that has been extensively discussed in \citet{dimatteo14}: stars born in the innermost galaxy regions (where the metallicity attains the highest values, according to the assumed initial conditions) remain mostly confined there, whilst stars coming from the outer regions of the bar and beyond (up to the bar OLR) are preferentially found at higher latitudes (see, for ex, Fig.~7 in \citet{dimatteo14}). Note that the presence of a small classical bulge ($\sim$ 10\%) does not have a significant impact on the metallicity map shown in Fig.~\ref{metmap}, since the stellar distribution in the boxy bulge is dominated, at all latitudes and longitudes, by stars originating in the disk (see, for example, top panel of Fig.~11 in \citet{dimatteo14}).\\

The ability of pure bar instability processes, originated in the thin disk, to reproduce the $global$ kinematic and chemical characteristics  of the MW's bulge may be considered sufficient to rule out the presence of other components in the inner regions of the Galaxy. In this respect, it seems there is no need to invoke the presence of a massive classical bulge or a thick disk to reproduce the main trends observed. It would be sufficient that at the time of bar formation, the MW thin disk had a steep radial  metallicity gradient accompanied by a positive radial gradient of $[\alpha/\rm{Fe}]$ values. If such initial conditions were in place, after the buckling instability, the metal poor, $\alpha-$enriched stars would have been preferentially mapped at high latitudes in the bulge, whilst the most metal rich and not $\alpha-$enhanced would have stayed preferentially confined closer to the Galaxy midplane, giving rise to the observed vertical metallicity and $[\alpha/\rm{Fe}]$ gradients. In \citet{dimatteo14}, we have already commented on the weakness of such a scenario: it would imply that the
typical metallicity of the MW disk at 4--5 kpc from the Galaxy center was between $-$1 and $-$1.5 dex and such low metallicities are never reached in the MW thin disk \citep[see, for example][]{fuhrmann99, bovy12, haywood13}, but  are  typical of the metal-poor tail of the  thick disk and of the MW halo \citep[see, among others, ][]{beers95, reddy08, nissen10}.
Here we are interested in exploring the consequences such a scenario would have on the spatial redistribution of stars and on their kinematics, as a function of their chemistry, and to show that the characteristics it would imply do not agree with what we know about the MW's bulge components.


\subsection{Failures of a pure thin-disk bar instability scenario}

The first consequence of a pure thin-disk bar instability scenario  for the MW's bulge is that \emph{all stars in the bulge fields should be part of the boxy/peanut-shaped structure}, independent on their metallicities. For example, as \citet{martinez13} point out, "the lower limit to the metallicity in the bulge fields would be set by the outer boundary of the part of the disk which participates in the instability".\\

\begin{figure*}
\centering
\includegraphics[width=0.85\textwidth,angle=0]{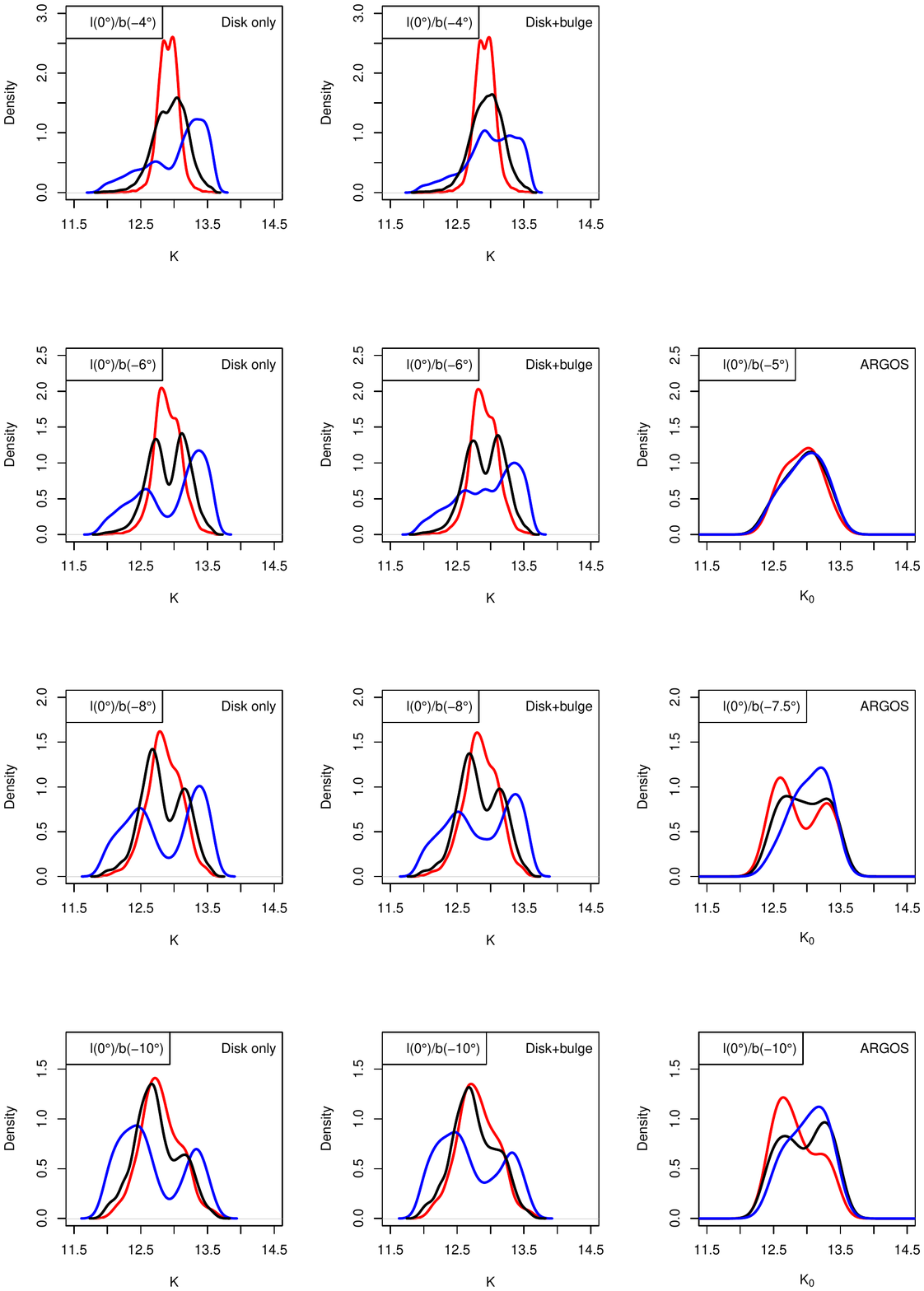}
\caption{\emph{(Left and middle columns): }K-band magnitude distributions of  red clump stars in the modeled galaxy at different latitudes along the bulge minor axis. An  initial radial metallicity profile [Fe/H]=0.5-0.4$R$ in the disk  is assumed, similarly to \citet{martinez13}. Three different metallicity bins are then shown: [Fe/H]$>0$ (red curve), -0.5$<$[Fe/H]$\le$0 (black curve), -1$<$[Fe/H]$\le$-0.5 (blue curve), similarly to populations A, B and C as defined by \citet{ness13a}. Disk stars only are shown in the left panel, while all (i.e. disk and classical bulge) stars are shown in the right panel. In all panels, stars with ($l,b$) in the interval ($l_0,b_0$)$\pm$($\Delta l, \Delta b$) are shown, with $\Delta l= \Delta b=1^\circ$ and  ($l_0,b_0$) as given in the top-left corner. \emph{(Right column): }$K_0$-band magnitude distributions of red clump stars at different latitudes along the bulge minor axis from the ARGOS survey. See \citet{ness13a} for details.} 
\label{peaks}
\end{figure*}

To quantify the implications of this scenario, in Fig.~\ref{peaks} we show the distribution of the apparent magnitude of red clump stars, as deduced from the N-body model, at different latitudes along the bulge minor axis. In producing this distribution, we have adopted an absolute magnitude for the clump stars of M$_K$=-1.61 \citep{alves00} which gives the minimum of the split red clump at K=12.9.

This distribution is particularly suitable to understand the underlying morphology. The density of a peanut-shaped bar indeed shows some depression in the center, on its minor axis, depression which is  more accentuated at higher latitudes. On the line of
sight, this produces a bimodal distribution, or a split, in the apparent K-magnitude of the red clump.

In Fig.~\ref{peaks}, the distribution of $K-$magnitudes at different latitudes is shown for stars in the metallicity ranges: [Fe/H]$>$0 dex, -0.5 dex $<$[Fe/H] $\le$0 dex, -1 dex $<$[Fe/H] $\le$ -0.5 dex, corresponding respectively to populations A, B and C, as defined by \citet{ness13a}. As in the previous section, the initial radial metallicity  gradient is $\alpha= -0.4$dex/kpc, and the classical bulge has a median metallicity of  -0.4 dex, and a dispersion of 0.3~dex.
Fig.~\ref{peaks} shows unequivocally  that if the MW's bulge was the result of a pure bar instability mechanism originated in the thin disk,  all stars in the bulge should be part of  the peanut structure, and in particular also those with -1 dex $<$[Fe/H] $\le$ -0.5 dex should show the split in the distribution of  $K-$magnitudes. But this is clearly in disagreement with the results presented by \citet{ness13a} -- and reported also in Fig.~\ref{peaks} --, who observationally found the split only for red clump stars of higher metallicities. Here thus lies the failure of the pure thin disk/bar instability model:  on the one hand, for the scenario to be successful, one would require metal-poor, $\alpha-$enriched stars to be present in the thin disk before the formation of the bar, in order to participate to the bar instability, be scattered at high vertical distances from the Galactic plane, and, as a consequence, constitute the low metallicity part of the metallicity distribution function; on the other hand, the natural and inevitable consequence of such a hypothesis is that those stars should be part of the peanut structure, while they are not, as, for example, the distribution of $K-$magnitudes deduced by the ARGOS survey shows. Note that even adding the contribution of a low mass classical bulge with B/D=0.1, which has been suggested as the upper limit of any classical bulge in the MW \citep{shen10, dimatteo14}, the result is unchanged (cfr left and right columns, Fig.~\ref{peaks}): all components, from A to C,  would be involved in the peanut, and show, as a consequence, a split in the distribution of red clump stars. 
In agreement with the findings of \citet{saha12, saha13}, such a low mass classical bulge, initially not-rotating, would acquire some angular momentum during the secular evolution (see Fig.~12 in \citet{dimatteo14}), but -- as we checked --  would not show any split in the K-magnitude distribution, indicating that it does not participate to the bar.  Overall, it would  leave only very weak signatures in the $K-$distribution, and, coupled with an initial  steep disk metallicity gradient, it would not erase the characteristic split of the peanut.

\begin{figure*}
\centering
\includegraphics[trim = 3cm 7cm 4cm 7.5cm, clip=TRUE,width=0.4\textwidth,angle=0]{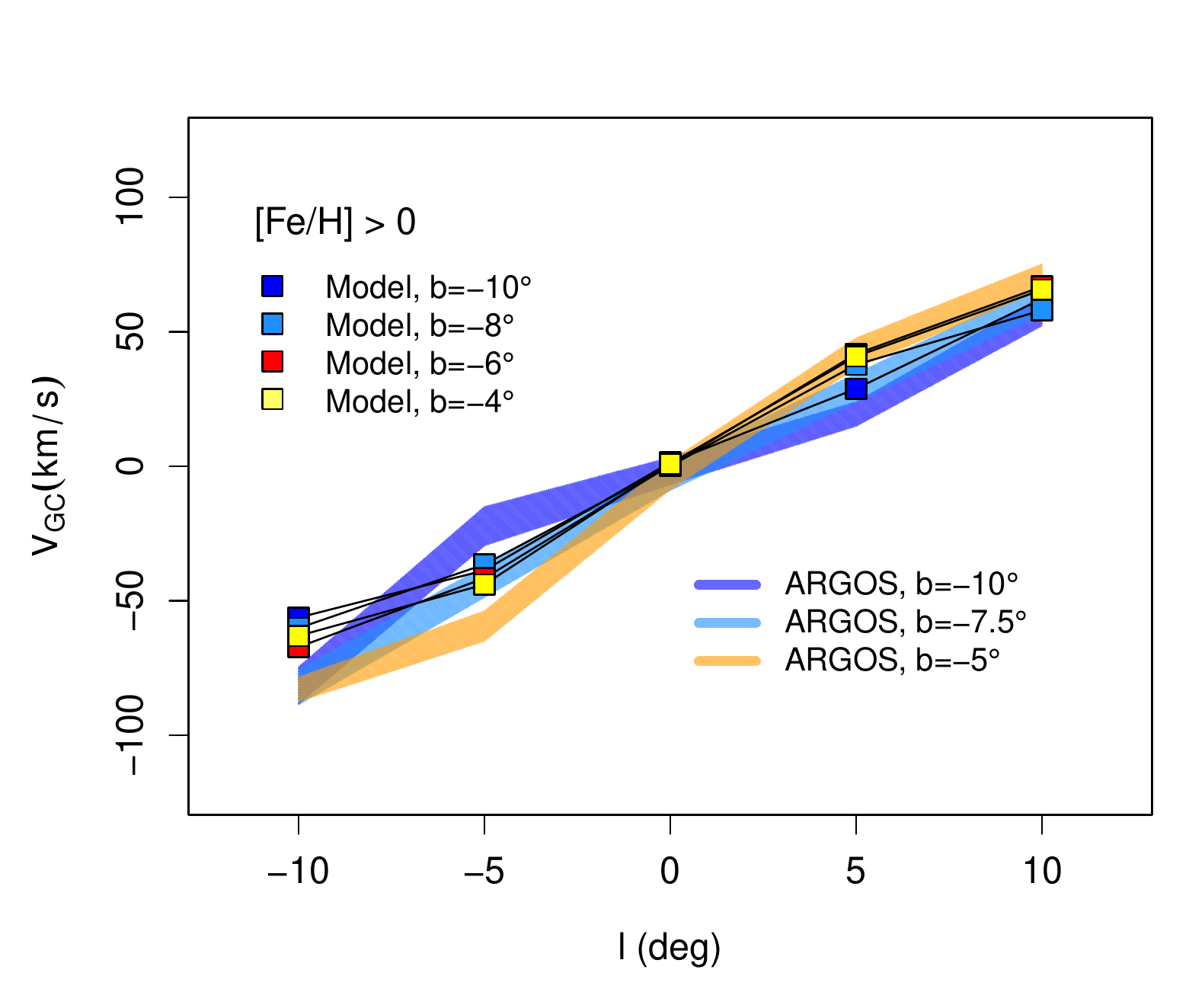}
\includegraphics[trim = 3cm 7cm 4cm 7.5cm, clip=TRUE,width=0.4\textwidth,angle=0]{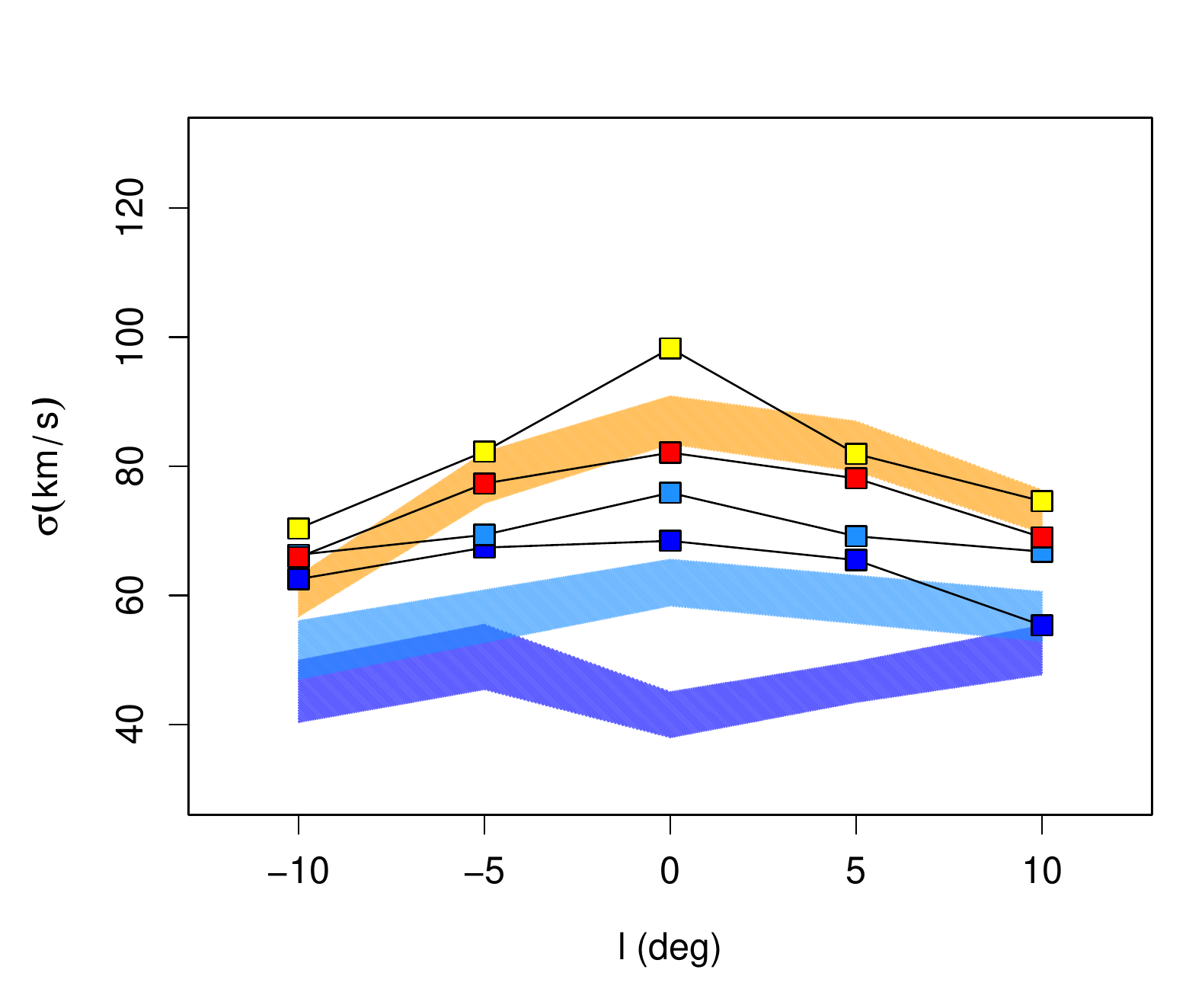}
\includegraphics[trim = 3cm 7cm 4cm 7.5cm, clip=TRUE,width=0.4\textwidth,angle=0]{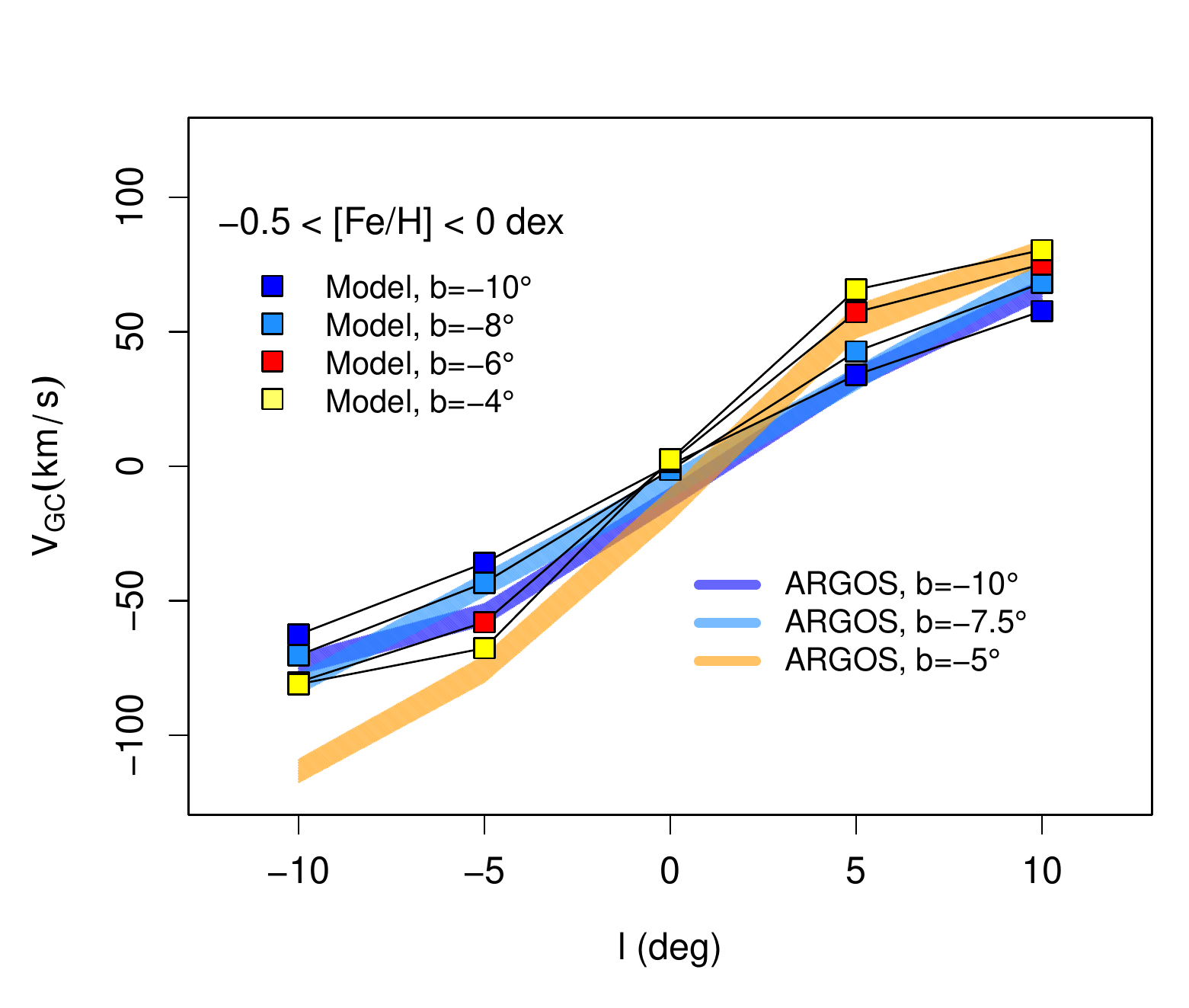}
\includegraphics[trim = 3cm 7cm 4cm 7.5cm, clip=TRUE,width=0.4\textwidth,angle=0]{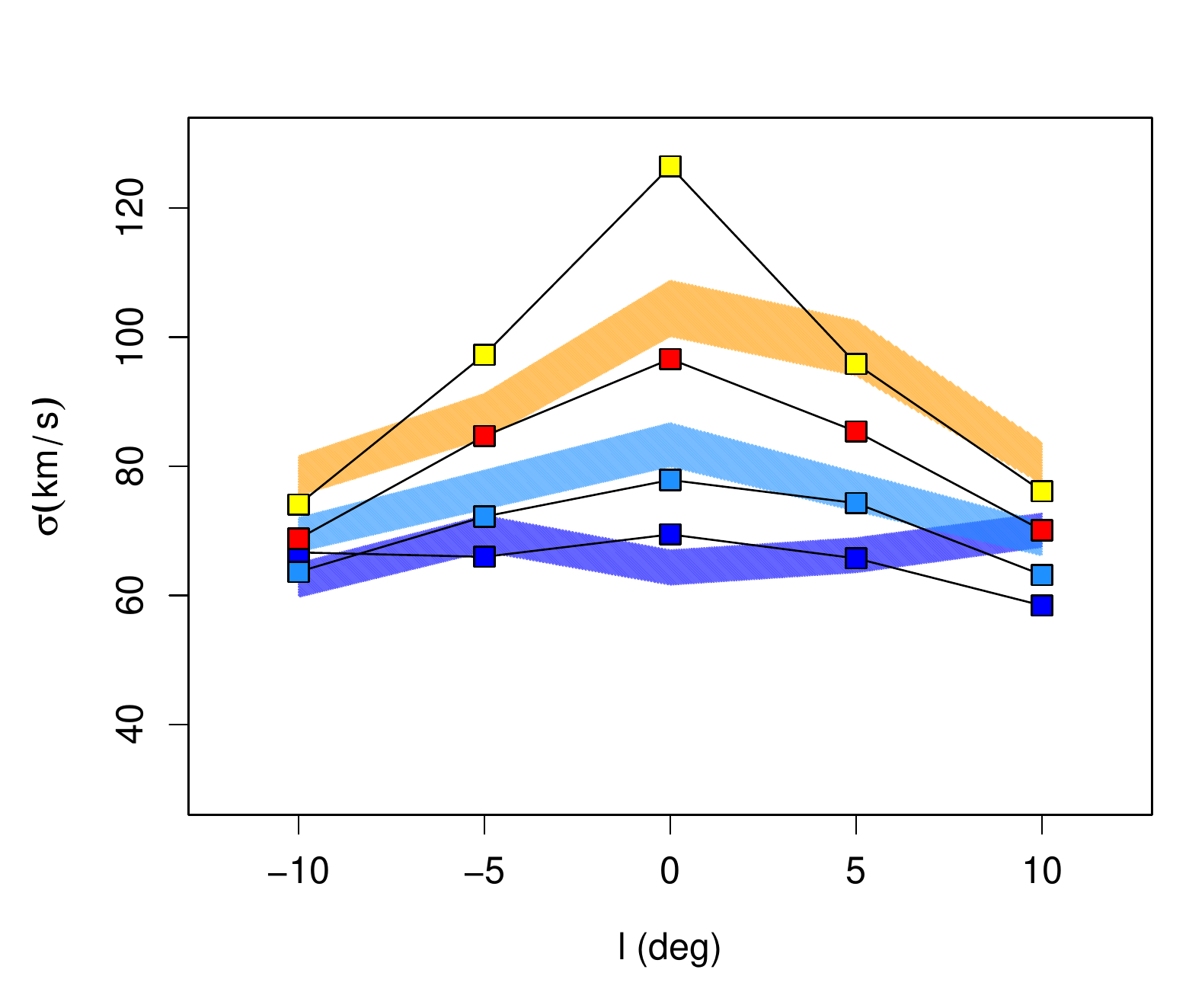}
\includegraphics[trim = 3cm 7cm 4cm 7.5cm, clip=TRUE,width=0.4\textwidth,angle=0]{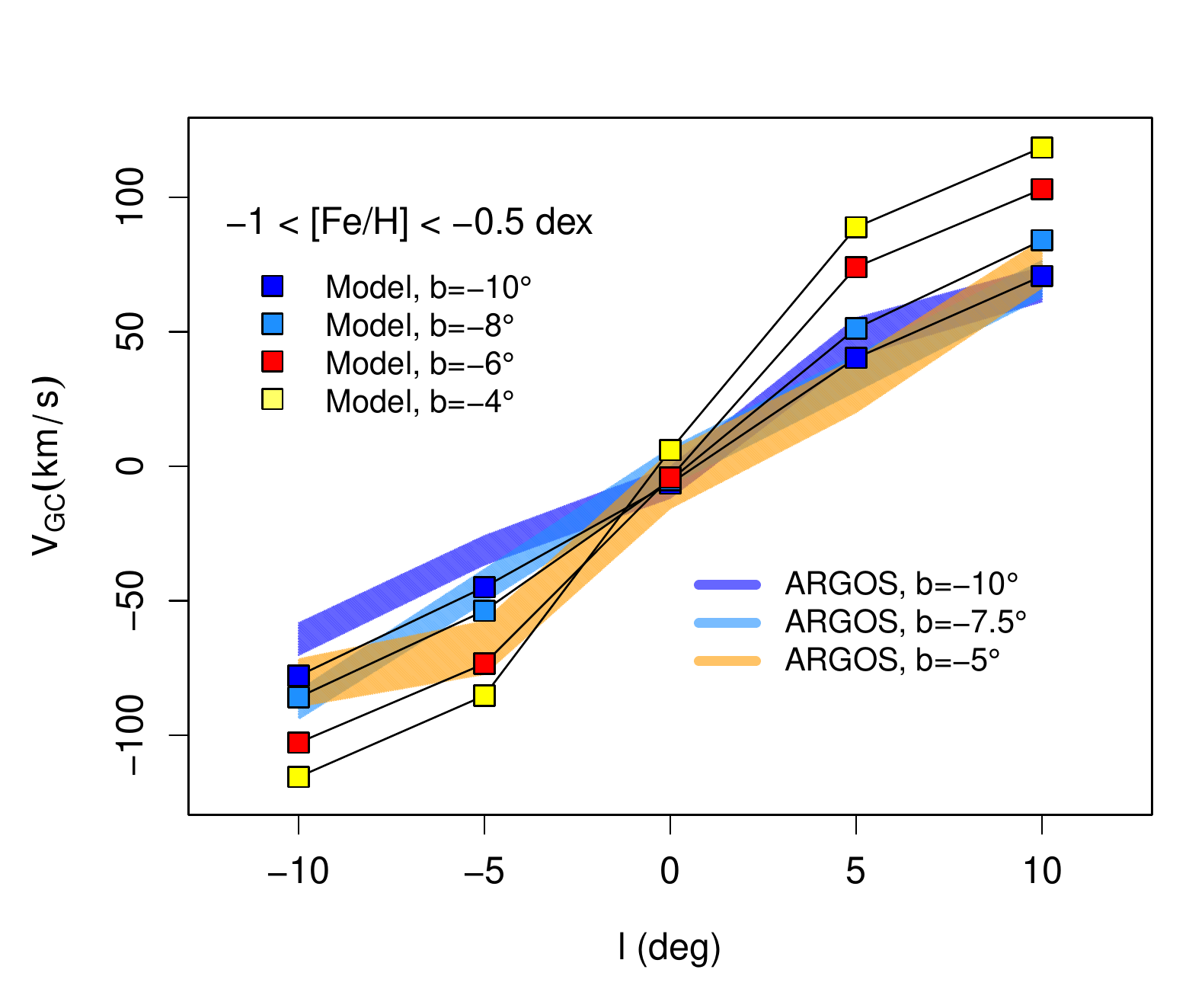}
\includegraphics[trim = 3cm 7cm 4cm 7.5cm, clip=TRUE,width=0.4\textwidth,angle=0]{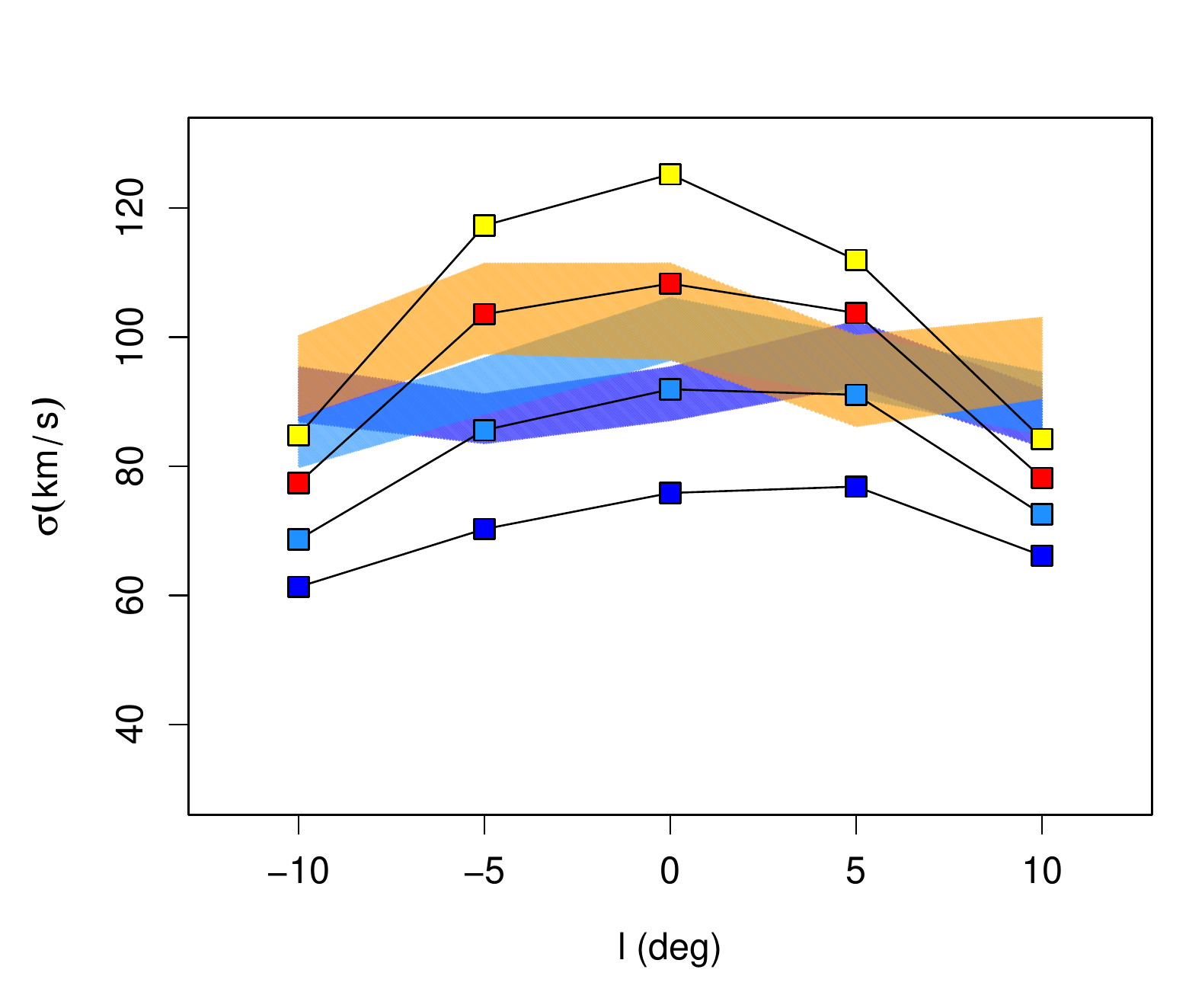}
\caption{Rotation curve \emph{(left panels)} and velocity dispersions \emph{(right panels)} a boxy bulge formed from a thin stellar disk (square symbols) compared to ARGOS data (colored curves). In the N-body model, only stars at $|x|\le 2.5$~kpc and $|y|\le 3$~kpc from the galaxy center are shown.
 An  initial radial metallicity profile [Fe/H]=0.5-0.4$R$ in the disk  is assumed, similarly to  \citet{martinez13}. Three different metallicity bins are shown, from top to bottom in decreasing [Fe/H], corresponding to the populations A, B and C as defined by \citet{ness13a}.  For each plot, four different latitudes are shown for the modeled galaxy: $b=-4^\circ$ (yellow squares), $b=-6^\circ$ (red squares), $b=-8^\circ$ (pale blue squares), $b=-10^\circ$ (dark blue squares). For comparison, ARGOS fields at  $b=-5^\circ$ (orange curve), $b=-7.5^\circ$ (pale blue curve), $b=-10^\circ$ (dark blue curve) for populations A, B and C are also given. The thickness of the curves corresponds to the $\pm 1 \sigma$ error in the observational data.}
\label{detkin}
\end{figure*}

The $K-$magnitude distribution of red clump stars is only an example of the discrepancy with observations that such a bulge formation scenario would imply. Also the $detailed$ kinematic characteristics of a bulge formed only via bar instability in a pre-existing thin stellar disk  would not agree with observations. To elucidate this point,  in Fig.~\ref{detkin} we show  the velocity curves and velocity dispersions profiles that stars in the three different metallicity bins formerly defined would have in such a scenario. Similarly to observations, which often lack proper motions measurements for bulge stars, only the radial component of the velocity is shown in this plot. For comparison, the velocities and velocity dispersions observed for populations A, B and C, as defined by \citet{ness13a}, are also shown.
If all stars in the bulge MDF had a thin disk origin, according to Fig.~\ref{detkin}, the following kinematic trends should be observed:
\begin{enumerate}
\item the velocity dispersion should increase with decreasing metallicity, in such a way that population C should be a warm replica of populations A and B, that is it should show velocity dispersions profiles similar to those of the stellar populations associated to the bar, but shifted to higher absolute values
\item the rotational velocity should increase with decreasing metallicity, continuously from super-solar metallicities to [Fe/H]= -1 dex.
\end{enumerate}
As explained in \citet{dimatteo14}, the trends at points 1. and 2. are the simple consequence of  the differential mapping of a thin disk into a boxy bulge,  the larger the birth radius of stars, the higher their  rotational velocity and  velocity dispersion.\\
In \citet{dimatteo14}, comparing these trends with the results by \citet{ness13b}, we have concluded that the MW's bulge population B (-0.5~dex $<$[Fe/H]$\le$0~dex) must have on average a more external disk origin than population A ([Fe/H]$>$0 dex). Indeed, according to the ARGOS observations reported in Fig.~\ref{detkin}, and to the discussion in \citet{ness13b},  population B shows a higher rotational velocity and similar velocity dispersions profiles (but with higher absolute values) than population A, as it would be the case if component B formed on average further out from the galaxy center than component A.\\
However, for the purpose of the present discussion,  we want to emphasize that points 1. and 2. also lead us to exclude the idea that the more metal poor population C  (-1~dex $<$[Fe/H]$\le$-0.5~dex) can have an origin in the old MW  thin disk. From \citet{martinez13} and Sect.~3.1, we have indeed learnt that a steep metallicity radial gradient in the thin disk is necessary before bar formation occurs, in order to reproduce a vertical gradient along the bulge minor axis, as currently observed.
Thus in this scenario, the metal poor population C would have been initially --i.e. before the onset of the bar instability -- a thin disk population located further out from the center than populations A and B. The subsequent bar vertical instability process would have scattered these stars to high distances from the plane, leading this component to have kinematic properties similar to those found in Fig.~\ref{detkin} for the most metal poor stars.  That is, in this scenario  C should show a rotational velocity greater than that of populations A and B, and its velocity dispersions would be a simple warm replica of those of the most metal rich populations. But this is excluded by the observational data  \citep{ness13b}, which shows velocity dispersions that are constant both with longitude and latitude, and a rotational velocity similar to that of population A and B.\\

Overall, our results point to the necessity of making use of both the kinematic and elemental abundance information from the various bulge
components to understand its origin. While this is obviously true, why is it that most analyses based on N-body simulations have not been
able to capture the kinematically warmer component that is apparent in the data? The reason is simple. 
Among the main components A, B and C found in the MW's bulge, most of the numerical models \citep{shen10, kunder12, ness13b, zoccali14}, including this work, have been essentially able to capture only component B.\\
B constitutes the backbone of the MW's bulge \citep[see][]{ness13a} and the bulge global kinematic trends are well represented by population B \citep[see Figs.3 and 6 in][]{ness13b}. Thus a pure thin disk N-body model which fits the global kinematics of the MW's bulge by means of a single stellar population is essentially fitting, and capturing,  the properties of population B only\footnote{In our model, this can be appreciated in Figs.~1 and 4:  from Fig.~4 it is evident that the modeled population B  is the only which satisfactory captures  the kinematics  of the corresponding observed stellar component (both components with [Fe/H] $>$ 0 and [Fe/H]$< -0.5$~dex do not capture the kinematics of the corresponding populations A and C observed by ARGOS); comparing Fig.~4 with Fig.~1, it is evident that the kinematics of modeled component B is representative  of the global bulge kinematics.}.
It is only by modeling both the peanut-shaped components A and B that N-body simulations can point out the existence of a kinematically warmer component like C. This because the inclusion of component A would result in a bulge with a global kinematics colder than what generally found in models which reproduce B only, and, as a consequence, it would be possible and even necessary to accommodate a kinematically warmer component to fit the global trends.


\section{Discussion: The Milky Way as a pure (thin+thick) disk galaxy}

If a pure thin disk/bar instability scenario for the MW's bulge can be rejected, does this imply that our Galaxy is not a pure disk galaxy, as it was suggested by \citet{shen10}? This question is tightly  related to this second one: What is the nature of the kinematically hotter component C found in the inner Galactic region, whose origin cannot be related to the thin disk?

The answer to this second question is still somewhat speculative at this stage, but it is worth suggesting that the growing evidence and
recent results should inspire us to critically revise our view of the populations and components of the MW.
From the work of \citet{bensby11} and \citet{bovy12} we learn that the $\alpha-$enhanced, metal-poor population of the Galaxy, coincident with the thick disk at the solar vicinity, has a scale length of about 1.8 kpc, that is a factor of about 2 less than previous estimates \citep[see for example][]{juric08}. This finding, coupled with the thin/thick disk local normalization, implies that the thick disk is under-represented at the solar vicinity, and it is rather mostly concentrated towards the inner Galactic regions, where it can become comparable to the thin disk mass \citep[see discussion in ][and also \citet{fuhrmann12}]{snaith14a}. 
The finding that the thick disk mass is comparable to that of the thin
disk is not only the result of the simple structural arguments given
previously, it finds a strong support, and independent confirmation,
in the recent work by \citet{snaith14a,snaith14b}. Snaith et al. modelled the
[alpha/Fe]-age relation recently discovered for disk stars in the solar
vicinity (Haywood et al. 2013) with a simple closed-box evolutionary
scenario. Snaith et al. showed that: i) the thick disk formed
during the most intense phase of star formation in the Galaxy; ii) this
phase, which lasted from about 13 to 8-9 Gyr ago, formed as many stars
as the subsequent, more quiescent, phase of star formation in the disk,
that proceeded from 8--9 Gyr ago to the present epoch during which the
thin disk was formed.
Note that the findings of \citet{snaith14a} are in substantial agreement with the independent work of \citet{vandokkum13}, who, on the basis of abundance matching techniques, followed the evolution of "MW-like"  progenitors from z=2.5 to the present epoch, concluding that these galaxies formed half of their stellar mass before redshift 1 \citep[see][and also \citet{lehnert14}]{snaith14a}. Their results indicate that before z=1, these galaxies were growing at all radii, with the inner regions growing at the same rate as the outer regions. As \citet{vandokkum13} point out, in these MW-analogues "we do not see high- density Ònaked bulgesÓ at z $\sim$ 2 around which disks gradually assembled. .... The evolution from z = 2.5 to z = 1 is strikingly uniform: the profiles are roughly parallel to one another, and rather than assembling only inside out the galaxies increase their mass at all radii. This is in marked contrast to more massive galaxies, which form their cores early and exclusively build up their outer parts over this redshift range." 
The interpretation of \citet{vandokkum13} results, in light of the works by \citet{snaith14a,snaith14b} and \citet{lehnert14}, strongly suggest that: 
\begin{itemize}
\item in these high redshift MW analogues we are witnessing the formation of  thick disks; 
\item these galaxies were not forming at those times any significant classical bulge. 
\end{itemize}
While we refer to \citet{lehnert14} for a detailed discussion on the properties of the MW at high redshift, here we emphasize that, 
 if confirmed, this formation and evolutionary scenario would  be substantially different from those suggesting that the old, metal-poor, $\alpha-$enhanced population of the MW's  bulge can be explained by a classical bulge or old spheroid \citep{babusiaux10, gonzalez11, hill11, uttenthaler12, zoccali14}. Recently, for example,  \citet{zoccali14}  suggested that the oldest component of MW's bulge resembles a low massive early-type galaxy, with properties similar to those of current early-type galaxies, as described by the SAURON and ATLAS3D samples \citep{bacon01,cappellari11}. Whilst it is true that an old thick disk can resemble for many aspects a fast rotating early-type system, the classical bulge/old spheroid scenario would imply that what we observe in the MW inner regions cannot be explained essentially on the basis of the known Galactic populations at the solar vicinity (i.e. thin and thick disks and only very marginally by a stellar halo), but rather requires to include an additional kinematically hot component, not negligible in terms of mass, in the central galactic regions. The disagreement between the two scenarios is thus much more than a simple semantic difference.
  Even the advocation \citep{zoccali14} of an instability mechanism via clumps formation and coalescence in the galaxy center \citep{noguchi99, immeli04, elmegreen08} to form the MW's bulge raises some doubts. The most recent models show that this evolutionary channel is inefficient in forming massive ($B/D> 10\%$) classical bulges in MW-like galaxies \citep{bournaud14}. Moreover, N-body simulations of classical bulge formation by clumpy instabilities show that these structures are usually slow rotators \citep[see][]{elmegreen08}, a property  hardly reconcilable with the fast rotation of population C \citep[see][]{ness13b}.  These considerations, coupled with the fact that the metallicity of component C is unlikely for a classical bulge of  $\sim 10^{10}M_{\odot}$ \citep[see discussion in][ and references therein]{dimatteo14} strongly suggest  the possibility that the MW's bulge is the result of the simple mapping of the Galactic (thin + thick) disk in the central regions of the Galaxy, with the kinematically coldest part of the disk  captured in the bar vertical instability (and thus mapped into populations A and B) and the kinematically hottest component forming population C \citep[see also the discussion in][]{ness13b}. Note that N-body simulations of boxy/peanut shaped bulges formed in galaxies containing both a thin and a thick stellar disk seem to support this scenario, by showing that the strength of the peanut shaped structure depends on the origin of the stars, with thin disk stars showing a more prominent peanut-shaped bar than stars originating in the thick disk \citep[see][their Fig.~10]{bekki11a}. These (thin+thick) disk models can also reproduce both  the observed cylindrical rotation and vertical metallicity gradient of the Galactic bulge reasonably well \citep{bekki11b}.


Even if further models are needed to test if this scenario  is able to reproduce the chemo-kinematic trends unraveled  by the most recent spectroscopic surveys of the Galactic bulge,  we emphasize that currently:
\begin{enumerate}
\item solar vicinity data  \citep[see][]{fuhrmann12, haywood13, snaith14a,snaith14b}), 
\item as well as Galactic kpc-scale observations \citep{bensby11, bovy12},  
\item the redshift evolution of MW-like analogues \citep{vandokkum13},
\end{enumerate}
\emph{all} seem to imply a substantial role of thick disks in the evolution of MW-like galaxies, a role still  underestimated in the current debate on  the nature of the  MW's bulge populations. The recent findings of strong similarities between the [$\alpha/$Fe] versus [Fe/H] trends of solar vicinity thick disk stars with the $\alpha-$enhanced, metal-poor population of the bulge \citep{melendez08, alves10,  bensby11, bensby13, gonzalez11}, if confirmed with larger high resolution spectroscopic samples, would be a strong additional support to this scenario. 

\section{Conclusions}

By analyzing a high resolution, N-body simulation of a bulge formed via a simple bar instability mechanism in a thin disk:
\begin{itemize}
\item We have shown that such a scenario is not compatible with the known structural and kinematic properties of the main populations of the Galactic bulge.
\item In particular, we emphasize that  \emph{global} kinematic and metallicity trends alone are not sufficient to constraint the MW's bulge formation scenario. It is only by coupling kinematic \emph{and} abundances information that N-body models are able to reject a pure thin disk/bar instability process to explain the formation and characteristics of the MW's bulge.
\item  Thus, in disagreement with recent suggestions, we conclude that the Milky Way bulge is not purely a  bar originated in a kinematically cold stellar disk and  seen edge-on. Its components did not all originate in the thin disk. 
\end{itemize}
 On the basis of a number of recent observational evidence, recalled in this paper, we suggest that the metal poor, $\alpha-$enhanced population which is present in the bulge, but which is not part of the peanut structure, is the same population known at the solar vicinity as  the old thick disk.

\section*{Acknowledgments}
The authors acknowledge the support of the French Agence Nationale de la Recherche (ANR) under contract ANR-10-BLAN-0508 (GalHis project). MN acknowledges funding 
from the European Research Council under the European UnionÕs  Seventh Framework Programme (FP 7) ERC Grant Agreement n. [321035].  Support for ONS was partially provided by NASA through the Hubble Space Telescope Archival Research grant HST-AR-12837.01-A from the Space Telescope Science Institute, which is operated by the Association of Universities for Research in Astronomy, Incorporated, under NASA contract NAS5-26555. The authors are grateful to N.~Stefanovitch for his technical help. PDM thanks F. Matteucci and C. Morossi for the organization of the stimulating conference on the "Formation and Evolution of the Galactic Bulge" (Sesto, Italy, January 2014), where many of the ideas presented in this paper took shape. PDM is grateful to F. Bournaud for an enriching discussion on the current state of  knowledge of clumpy disk galaxies.
We thank the referee for a very constructive report, which greatly helped us to clarify our results.

\end{document}